\documentclass[conference, letterpaper]{IEEEtran}
\IEEEoverridecommandlockouts
\usepackage{amsmath,amssymb,amsfonts}
\usepackage{algorithm, algorithmic}
\usepackage{graphicx}
\usepackage{textcomp}
\usepackage{xcolor}

\usepackage{subcaption}
\usepackage{amsmath,amsthm,amssymb}
\usepackage{floatflt}  
\usepackage[hyphens]{url}
\usepackage{hyperref}

\usepackage[left=1.62cm,right=1.62cm,top=1.8cm]{geometry}

\usepackage{comment}
\usepackage{flushend}

\begin{document}

\title{Securing Blockchain-based IoT Systems with Physical Unclonable Functions and Zero-Knowledge Proofs
}

\author{
\IEEEauthorblockN{1\textsuperscript{st} Daniel Commey}
\IEEEauthorblockA{\textit{Multidisciplinary Engineering} \\
\textit{Texas A\&M University}\\
Texas, USA\\
dcommey@tamu.edu
}
\and
\IEEEauthorblockN{2\textsuperscript{nd} Sena Hounsinou}
\IEEEauthorblockA{\textit{Computer Science \& Cybersecurity} \\
\textit{Metro State University}\\
Minnesota, USA\\
sena.houeto@metrostate.edu
}
\and
\IEEEauthorblockN{3\textsuperscript{rd} Garth V. Crosby}
\IEEEauthorblockA{\textit{Engineering Technology \& Industrial Distribution} \\
\textit{Texas A\&M University}\\
Texas, USA\\
gvcrosby@tamu.edu
}
}

\maketitle

\begin{abstract}
This paper presents a framework for securing blockchain-based IoT systems by integrating Physical Unclonable Functions (PUFs) and Zero-Knowledge Proofs (ZKPs) within a Hyperledger Fabric environment. The proposed framework leverages PUFs for unique device identification and ZKPs for privacy-preserving authentication and transaction processing. Experimental results demonstrate the framework's feasibility, performance, and security against various attacks. This framework provides a comprehensive solution for addressing the security challenges in blockchain-based IoT systems.
\end{abstract}

\begin{IEEEkeywords}
Blockchain, Internet of Things, Zero-Knowledge Proofs, Physical Unclonable Functions, Hyperledger Fabric, Device Security, Data Integrity
\end{IEEEkeywords}

\section{Introduction}
\label{sec:introduction}

The rapid growth of the Internet of Things (IoT) devices has revolutionized various sectors by enabling smart environments through interconnected devices that collect and exchange data. However, this expansion has also introduced significant security concerns, particularly in safeguarding communication and data integrity among IoT devices \cite{ramezan_zk-iot_2024}. Blockchain technology has emerged as a promising solution to enhance IoT security by ensuring data integrity, transparency, and decentralization \cite{wang_blockchain-based_2022}. Yet, integrating blockchain with IoT presents new challenges, especially in terms of scalable and efficient device authentication and data privacy \cite{zhong_-demand_2023}.

Physical Unclonable Functions (PUFs) offer a unique approach to device security by providing each IoT device with a distinct, inseparable identity derived from its inherent physical variations \cite{zhong_-demand_2023}. This feature makes PUFs an ideal candidate for enhancing device authentication processes. However, to fully leverage PUFs within blockchain-based IoT (BIoT) systems, there is a need for a framework that can accommodate the PUFs' unique characteristics without compromising privacy or scalability \cite{ramezan_zk-iot_2024}.

Zero-knowledge proofs (ZKPs) have emerged as a complementary technology to address these challenges. ZKPs allow one party to prove to another that a statement is true without revealing any information beyond the validity of the statement itself \cite{chi_privacy-preserving_2023}. Integrating ZKPs with PUFs within a blockchain environment provides a dual-layered security mechanism. This integration strengthens the authentication process and ensures that transactional data among IoT devices remain private and secure \cite{li_aggregated_2023}.

Several recent studies have explored the integration of PUFs, ZKPs, and blockchain technologies to enhance the security and privacy of IoT systems. However, these works often focus on specific use cases or lack a comprehensive framework that addresses the challenges of device authentication, data privacy, and scalability in a unified manner. For instance, Zhong et al. \cite{zhong_-demand_2023} proposed a PUF-based authentication protocol for IoT devices but did not consider the integration with blockchain technology. Ramezan and Meamari \cite{ramezan_zk-iot_2024} introduced a ZKP-based framework for secure IoT firmware updates using blockchain, but their approach does not incorporate PUFs for device authentication. While these studies have made valuable contributions, there remains a need for a holistic framework that seamlessly integrates PUFs, ZKPs, and blockchain to provide a robust, scalable, and privacy-preserving solution for BIoT systems.

\subsection{Contributions}
This paper introduces a framework that integrates PUFs and ZKPs to secure BIoT systems. Our proposed framework combines IoT devices equipped with PUFs, a blockchain network, and a ZKP-based authentication and communication protocol. This unique integration enables secure and privacy-preserving device authentication, ensures data integrity and confidentiality, and provides a scalable and tamper-evident ledger for recording IoT transactions. By leveraging the inherent security features of PUFs and the privacy-preserving capabilities of ZKPs within a permissioned blockchain environment, our framework addresses the key challenges of device authentication, data privacy, and scalability in BIoT systems.

Our contributions are summarized as follows:
\begin{enumerate}
\item We propose a unique integration of PUFs, ZKPs, and blockchain technologies tailored for securing IoT systems, addressing specific challenges related to device authentication, data privacy, and scalability.

\item Our framework leverages the inherent security features of PUFs to provide each IoT device with a unique, unclonable identity, enhancing the device authentication process and mitigating risks associated with device impersonation and cloning.

\item We employ ZKPs to enable privacy-preserving authentication and data integrity, ensuring that device identities and transactional data can be verified without revealing sensitive information.

\item Our model utilizes the Hyperledger Fabric blockchain platform to provide an immutable and tamper-evident ledger for recording IoT transactions, ensuring data integrity and traceability.

\item We present a comprehensive security analysis of our proposed framework, examining its resilience against various threats, including impersonation attacks, man-in-the-middle attacks, and data tampering attempts, drawing from the threat models and security analyses presented in related works.

\item We demonstrate the feasibility and effectiveness of our model through extensive simulations, providing insights into its practical implementation and performance in real-world IoT scenarios, building upon the simulation methodologies employed in previous studies.
\end{enumerate}

The rest of the paper is organized as follows. Section~\ref{sec:related_works} reviews related works, Section~\ref{sec:preliminaries} introduces essential concepts, Section~\ref{sec:proposed_model} presents our proposed framework, Section~\ref{sec:implementation_results} evaluates the framework's performance, and Section~\ref{sec:conclusion} concludes the paper and discusses future research directions.

\section{Related Works}
\label{sec:related_works}

The integration of blockchain technology and IoT has gained significant attention in recent years, with numerous studies exploring the potential of this combination to enhance security and privacy in IoT systems. Wang et al. \cite{wang_blockchain-based_2022} proposed a block verification mechanism based on zero-knowledge proofs to improve the efficiency and privacy of blockchain data verification in IoT environments. While their approach optimizes verification speed without compromising user privacy, it does not address the challenge of secure device authentication, which is crucial in IoT systems.

Zhong et al. \cite{zhong_-demand_2023} introduced an efficient and secure on-demand communication protocol using zero-knowledge proofs to secure IoT devices against hardware attacks. Although their protocol enables edge devices to authenticate themselves to a central server without revealing device-specific secrets, it does not leverage the decentralized nature of blockchain technology to enhance the overall security and scalability of the IoT ecosystem.

Ramezan and Meamari \cite{ramezan_zk-iot_2024} proposed the zk-IoT framework, which integrates zero-knowledge proofs with blockchain technology to secure IoT ecosystems. While their framework ensures the integrity of firmware execution and data processing in potentially compromised IoT devices, it does not incorporate PUFs for robust device authentication, which is a key feature of our proposed framework.

Miyamae et al. \cite{miyamae_zgridbc_2023} introduced ZGridBC, a scalable and privacy-enhanced blockchain platform for tracking electricity production and consumption in renewable energy facilities. Although ZGridBC leverages zero-knowledge proofs to aggregate massive data and ensure privacy and scalability, it is tailored specifically for the energy sector and may not be easily adaptable to other IoT domains.

Chi et al. \cite{chi_privacy-preserving_2023} developed the Blockchain Designated Verifier Proof (BDVP) scheme, enabling non-transferable zero-knowledge proofs to maintain privacy in blockchain transactions. While the BDVP scheme ensures that only authorized verifiers can validate the truthfulness of a claim without compromising the prover's privacy, it does not address the challenge of secure device authentication in IoT systems.

Jeong et al. \cite{jeong_azeroth_2023} proposed Azeroth, a zero-knowledge proof-based framework that verifies encrypted transactions while ensuring privacy. Although Azeroth integrates zk-SNARKs with encrypted transactions and incorporates features for authorized auditors, it does not leverage the unique properties of PUFs for device authentication and is not specifically designed for IoT environments.

In the context of vehicular ad hoc networks (VANETs), Kalmykov et al. \cite{kalmykov_using_2023} and Chistousov et al. \cite{chistousov_adaptive_2022} proposed adaptive zero-knowledge authentication protocols that enhance privacy while optimizing time costs and data exchange volumes based on traffic density. While these protocols show significant improvements in vehicle authentication and privacy, they are limited to the specific domain of VANETs and do not address the broader challenges of securing blockchain-based IoT systems.

Li et al. \cite{li_aggregated_2023} proposed an authentication system for autonomous truck platooning that combines aggregated BLS signatures with blockchain technology. While their system demonstrates low latency, high throughput, and constant verification time, their approach does not utilize well-established zero-knowledge proof schemes such as zk-SNARKs, zk-STARKs, or Bulletproofs. Instead, they adapt the BLS signature scheme to achieve some properties of zero-knowledge proofs. As a result, their performance metrics may not be directly comparable to those of systems built on proven ZKP schemes, which have undergone rigorous analysis and have been shown to provide strong security and privacy guarantees.

In contrast to these studies, our proposed framework offers a comprehensive solution that integrates PUFs and well-established zk-SNARK proofs within a blockchain environment. Our framework leverages the unique properties of PUFs for robust device authentication and utilizes zk-SNARKs for privacy-preserving transaction processing, addressing the key challenges of security, privacy, and scalability in blockchain-based IoT systems. Moreover, our framework is designed to be modular and adaptable, making it suitable for a wide range of IoT applications beyond specific domains like energy or transportation. These distinctive features, along with our detailed security analysis and performance evaluation, demonstrate the novelty and value of our contribution to the field of securing blockchain-based IoT systems.

\section{Preliminaries}
\label{sec:preliminaries}

In this section, we introduce the key concepts and technologies that form the foundation of our proposed framework for securing blockchain-based IoT systems. We provide a detailed overview of Hyperledger Fabric, Zero-Knowledge Proofs (ZKPs), and Physical Unclonable Functions (PUFs), along with their mathematical formulations and relevant applications in the context of IoT security.

\subsection{Hyperledger Fabric}
\label{subsec:hyperledger_fabric}
Hyperledger Fabric is an open-source, permissioned blockchain platform for developing enterprise-grade applications \cite{androulaki_hyperledger_2018}. It offers a modular architecture that supports the development of blockchain solutions with a focus on scalability, flexibility, and confidentiality. Fabric utilizes a unique execute-order-validate approach, which separates the transaction execution from the ordering and validation phases, enabling parallel processing and improved performance \cite{gorenflo_fastfabric_2020}.

One of the critical features of Hyperledger Fabric is its support for smart contracts, referred to as "chaincodes." Chaincodes are self-executing contracts that encapsulate the business logic of an application and are triggered by transaction proposals \cite{dhillon_hyperledger_2017, noauthor_smart_2024}. The chaincode execution model in Fabric can be formally represented as follows:

\begin{equation}
\label{eq:chaincode}
s' = f(s, t)
\end{equation}

where $s$ represents the current state of the ledger, $t$ is the transaction proposal, $f$ is the chaincode function, and $s'$ is the resulting state after executing the chaincode.

Fabric also incorporates a unique membership service provider (MSP) component, which manages the identities and permissions of network participants \cite{noauthor_membership_2024}. The MSP uses public key infrastructure (PKI) and certificate authorities (CAs) to issue and validate digital certificates, ensuring secure communication and authentication within the network.

\subsection{Zero-Knowledge Proofs (ZKPs)}
\label{subsec:zero_knowledge_proofs}
Zero-knowledge proofs (ZKPs) are cryptographic techniques that allow one party (the prover) to prove to another party (the verifier) that a given statement is true without revealing any additional information beyond the validity of the statement itself \cite{goldwasser_knowledge_2019}. ZKPs have gained significant attention in recent years due to their potential to enhance privacy and security in various applications, including blockchain systems \cite{kosba_hawk_2016}.

A ZKP protocol must satisfy three key properties \cite{blum_non-interactive_2019}:

\begin{enumerate}
\item \textbf{Completeness}: If the statement is true, an honest prover can convince an honest verifier of its validity.
\item \textbf{Soundness}: If the statement is false, no cheating prover can convince an honest verifier that it is true, except with negligible probability.
\item \textbf{Zero-Knowledge}: The verifier learns nothing beyond the validity of the statement.
\end{enumerate}

One of the most widely used ZKP constructions is the Zero-Knowledge Succinct Non-Interactive Argument of Knowledge (zk-SNARK) \cite{ben-sasson_succinct_2014}. zk-SNARKs enable the creation of succinct and non-interactive proofs, making them well-suited for use in blockchain applications. The general structure of a zk-SNARK can be described as follows:

\begin{equation}
\label{eq:zksnark}
\pi = \text{Prove}(x, w)
\end{equation}

where $x$ is the public input, $w$ is the private witness, and $\pi$ is the resulting proof. The verifier can then check the validity of the proof using a verification function:

\begin{equation}
\label{eq:zksnark_verify}
\text{Verify}(x, \pi) = \begin{cases}
1 & \text{if } \pi \text{ is a valid proof for } x \\
0 & \text{otherwise}
\end{cases}
\end{equation}

zk-SNARKs have been successfully implemented in various blockchain platforms, such as Zcash \cite{ben-sasson_succinct_2014} and Ethereum \cite{buterin_ethereum_2014}, to enable privacy-preserving transactions and smart contract execution.

\subsection{Physical Unclonable Functions (PUFs)}
\label{subsec:pufs}
Physical Unclonable Functions (PUFs) are hardware-based security primitives that exploit the inherent manufacturing variations in integrated circuits to generate unique and unclonable device identifiers \cite{gassend_controlled_2002}. PUFs provide a cost-effective and tamper-evident solution for device authentication and key generation in IoT systems \cite{herder_physical_2014}.

A PUF can be modeled as a function that maps a set of challenges ($C$) to a set of responses ($R$):

\begin{equation}
\label{eq:puf}
PUF: C \rightarrow R
\end{equation}

The unique physical characteristics of each device ensure that the mapping between challenges and responses is device-specific and practically impossible to clone or predict. This property makes PUFs an attractive option for secure device authentication and key generation in IoT systems \cite{delvaux_machine-learning_2019}.

Several types of PUFs have been proposed in the literature, including delay-based PUFs (e.g., Arbiter PUF \cite{lee_technique_2004}), memory-based PUFs (e.g., SRAM PUF \cite{guajardo_physical_2007}), and hybrid PUFs (e.g., Bistable Ring PUF \cite{chen_bistable_2011}). Each type of PUF exploits different physical properties and has advantages and limitations regarding reliability, uniqueness, and security \cite{maes_physically_2013}.

PUFs have been successfully integrated with blockchain systems to enhance the security and privacy of IoT devices. For example, Xu et al. \cite{xu_coc_2017} proposed a blockchain-based IoT device authentication scheme using PUFs, while Liu et al. \cite{liu_backslashmathsf_2019} combined PUFs with smart contracts to enable secure and decentralized access control in IoT systems.

\section{Proposed Model}
\label{sec:proposed_model}
This section presents our framework for securing blockchain-based IoT systems by integrating Physical Unclonable Functions (PUFs) and Zero-Knowledge Proofs (ZKPs) within a Hyperledger Fabric environment. The proposed model aims to address the challenges of device authentication, data privacy, and scalability in IoT systems while leveraging the unique properties of PUFs and ZKPs. We describe the framework's architecture, authentication mechanism, security analysis, and the necessary mathematical formulations, proofs, and algorithms.

\subsection{System Architecture}
\label{subsec:system_architecture}
\begin{figure}
\centering
\includegraphics[width=\linewidth]{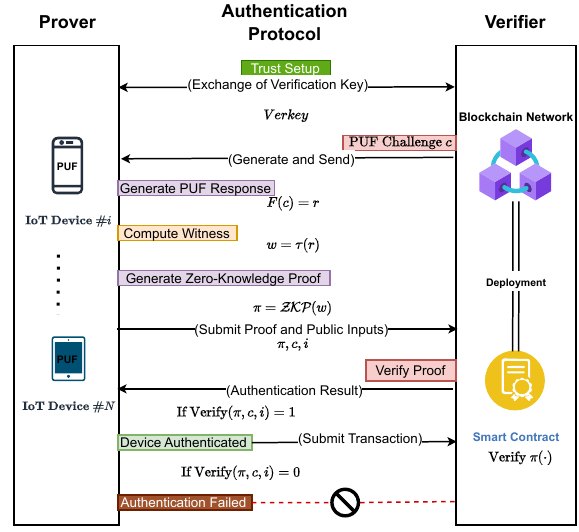}
\caption{System architecture of the proposed framework for securing blockchain-based IoT systems using PUFs and ZKPs.}
\label{fig:architecture}
\end{figure}

The proposed system architecture comprises three main components: IoT devices equipped with PUFs, a Hyperledger Fabric blockchain network, and a ZKP-based authentication and communication protocol. Figure \ref{fig:architecture} illustrates the overall architecture of the proposed framework and the interactions between its components.

Each IoT device has a PUF that generates a unique and unclonable device identifier based on the device's physical characteristics (see Section \ref{subsec:pufs}).

The IoT devices interact with the Hyperledger Fabric blockchain network through the ZKP-based authentication and communication protocol. The protocol enables secure and privacy-preserving authentication of IoT devices and ensures the integrity and confidentiality of the transmitted data.

The Hyperledger Fabric blockchain network serves as the system's backbone, providing a decentralized and tamper-evident ledger for storing device identities, authentication records, and transaction data. The blockchain network consists of nodes, including peers, orderers, and certificate authorities (CAs). Peers are responsible for maintaining the ledger and executing smart contracts, while orderers ensure the consistency and order of transactions. CAs manage the issuance and revocation of digital certificates used for authentication and secure communication within the network.

The ZKP-based authentication and communication protocol leverages zk-SNARKs to generate and verify proof of device authenticity and data integrity without revealing sensitive information to unauthorized parties. The protocol consists of three main phases: setup, authentication, and transaction processing. In the setup phase, the necessary cryptographic parameters are generated. The authentication phase allows IoT devices to prove their identity using their PUF-derived secrets and zk-SNARKs. The transaction processing phase enables secure and private execution of transactions between authenticated IoT devices.

\subsection{zk-SNARK Scheme Details}
\label{subsec:zk_snark_details}

The ZKP-based authentication protocol and secure transaction processing in our proposed framework rely on the use of zk-SNARKs (see Section \ref{subsec:zero_knowledge_proofs}).

In our proposed framework, we employ the Groth16 zk-SNARK scheme \cite{groth_size_2016}, which offers constant-size proofs and efficient verification. The specific implementation details of the Groth16 scheme in our framework are as follows:

Let $\mathbb{G}_1$, $\mathbb{G}_2$, and $\mathbb{G}_T$ be cyclic groups of prime order $p$ with generators $g_1$, $g_2$, and $g_T$, respectively. Let $e: \mathbb{G}_1 \times \mathbb{G}_2 \rightarrow \mathbb{G}_T$ be a bilinear pairing, which satisfies the following properties:

\begin{itemize}
\item Bilinearity: $\forall a,b \in \mathbb{Z}_p, e(g_1^a, g_2^b) = e(g_1, g_2)^{ab}$
\item Non-degeneracy: $e(g_1, g_2) \neq 1$
\item Computability: $e$ is efficiently computable
\end{itemize}

The zk-SNARK scheme consists of three main phases: setup, proving, and verification.

\textbf{Setup Phase:}
In the setup phase, the following public parameters are generated:
\begin{itemize}
\item A random value $\alpha \xleftarrow{\$} \mathbb{Z}_p$ is chosen, and the public key $pk$ is set as $pk \leftarrow g_2^{\alpha}$.
\item A hash function $H: \{0,1\}^* \rightarrow \mathbb{G}_1$ is selected. $H$ is modeled as a random oracle.
\end{itemize}

\textbf{Authentication Proving Phase:}
In the authentication proving phase, the prover (IoT device) generates a zk-SNARK proof $\pi_i$ for proving knowledge of the PUF response $R_i$ and private key $sk_i$ by performing the following steps:
\begin{align}
r &\xleftarrow{\$} \mathbb{Z}_p \\
S &\leftarrow H(R_i || sk_i)^r \\
U &\leftarrow g_1^r \\
h &\leftarrow H(S || U) \\
V &\leftarrow (H(R_i || sk_i) \cdot g_1^h)^r \\
\pi_i &\leftarrow (S, U, V)
\end{align}

where $r$ is a random value chosen from $\mathbb{Z}_p$, $S$ is a commitment to the PUF response $R_i$ and private key $sk_i$, $U$ is a commitment to the randomness $r$, $h$ is a challenge computed using the hash function $H$, and $V$ is a response to the challenge.

\textbf{Authentication Verification Phase:}
In the authentication verification phase, the verifier checks the validity of the zk-SNARK proof $\pi_i = (S, U, V)$ by verifying the equality of the bilinear pairings:
\begin{align}
h &\leftarrow H(S || U) \\
\text{auth\_result} &\leftarrow \begin{cases}
1 & \text{if } e(S, g_2) \cdot e(U, pk)^h = e(V, g_2) \\
0 & \text{otherwise}
\end{cases}
\end{align}

The verification process checks if the response $V$ is correctly computed using the commitments $S$ and $U$, and the public key $pk$.

\textbf{Transaction Proving Phase:}
In the transaction proving phase, the prover (IoT device) generates a zk-SNARK proof $\pi_{T_i}$ for a transaction $T_i$ by performing the following steps:
\begin{align}
r &\xleftarrow{\$} \mathbb{Z}_p \\
S &\leftarrow H(T_i)^r \\
U &\leftarrow g_1^r \\
h &\leftarrow H(S || U) \\
V &\leftarrow (H(T_i) \cdot g_1^h)^r \\
\pi_{T_i} &\leftarrow (S, U, V)
\end{align}

where $T_i$ is the transaction data, and the other variables are computed similarly to the authentication proving phase.

\textbf{Transaction Verification Phase:}
In the transaction verification phase, the verifier (smart contract) checks the validity of the zk-SNARK proof $\pi_{T_i} = (S, U, V)$ by verifying the equality of the bilinear pairings:
\begin{align}
h &\leftarrow H(S || U) \\
\text{proof\_result} &\leftarrow \begin{cases}
1 & \text{if } e(S, g_2) \cdot e(U, pk)^h = e(V, g_2) \\
0 & \text{otherwise}
\end{cases}
\end{align}

If the $\text{proof\_result}$ is 1, the transaction is considered valid and can be executed by the smart contract. Otherwise, the transaction is rejected.

\subsection{PUF-based Device Registration}
The PUF-based device registration process ensures that each IoT device has a unique and unclonable identifier derived from its physical characteristics. Let $\mathcal{D}$ be the set of all IoT devices, and let $D_i \in \mathcal{D}$ be an IoT device with a PUF $P_i$. The PUF $P_i$ can be modeled as a function that maps a set of challenges $C_i$ to a set of responses $R_i$:

\begin{equation}
P_i: C_i \rightarrow R_i
\end{equation}

The device registration process is described in Algorithm \ref{alg:device_registration}.

\begin{algorithm}
\caption{PUF-based Device Registration}
\label{alg:device_registration}
\begin{algorithmic}[1]
\renewcommand{\algorithmicrequire}{\textbf{Input:}}
\renewcommand{\algorithmicensure}{\textbf{Output:}}
\REQUIRE IoT device $D_i$ with PUF $P_i$, Hyperledger Fabric blockchain $B$, Certificate Authority $CA$
\ENSURE Unique identifier $ID_i$ and cryptographic keys $K_i$ for device $D_i$

\STATE \textit{Initialization}:

\STATE Generate a set of challenges $C_i \xleftarrow{\$} \{0,1\}^n$ for the PUF $P_i$

\STATE Obtain the corresponding set of responses $R_i \leftarrow P_i(C_i)$

\STATE Generate a public-private key pair $(pk_i, sk_i) \xleftarrow{\$} \mathcal{K}$ for device $D_i$

\STATE Create a unique device identifier $ID_i \leftarrow \text{Hash}(pk_i || R_i)$

\STATE Register the device identifier $ID_i$ and public key $pk_i$ with the $CA$

\STATE Store the device identifier $ID_i$, public key $pk_i$, and PUF challenges $C_i$ on the blockchain $B$

\STATE Set the device's cryptographic keys $K_i \leftarrow (pk_i, sk_i)$

\RETURN $ID_i, K_i$
\end{algorithmic}
\end{algorithm}

In Algorithm \ref{alg:device_registration}, ${0,1}^n$ represents the set of all binary strings of length $n$, and $\mathcal{K}$ represents the key space for the public-private key pair. The device identifier $ID_i$ is created by hashing the concatenation of the public key $pk_i$ and the PUF responses $R_i$ using a secure hash function $\text{Hash}(\cdot)$.

The PUF-based device registration process ensures the uniqueness and unclonability of the device identifier, as stated in the following theorem:

\textbf{Theorem 1.} \textit{The PUF-based device registration process ensures that each IoT device has a unique and unclonable identifier.}

\textit{Proof sketch.} The uniqueness of the device identifier $ID_i$ is guaranteed by the uniqueness of the PUF responses $R_i$, which are derived from the device's inherent physical variations. The unclonability of $ID_i$ is ensured by the unclonability of the PUF itself, as it is infeasible to create a physical replica of the PUF that yields the same challenge-response behavior. $\square$

\subsection{ZKP-based Authentication Protocol}
The ZKP-based authentication protocol enables secure and privacy-preserving authentication of IoT devices without revealing sensitive information. The protocol leverages zk-SNARKs to generate and verify proofs of device authenticity. The specific zk-SNARK construction used in this framework is the Groth16 scheme \cite{groth_size_2016}, which offers constant-size proofs and efficient verification.

The authentication protocol is described in Algorithm \ref{alg:authentication_protocol}.

\begin{algorithm}
\caption{ZKP-based Authentication Protocol}
\label{alg:authentication_protocol}
\begin{algorithmic}[1]
\renewcommand{\algorithmicrequire}{\textbf{Input:}}
\renewcommand{\algorithmicensure}{\textbf{Output:}}
\REQUIRE IoT device $D_i$ with identifier $ID_i$, cryptographic keys $K_i$, and PUF $P_i$, Hyperledger Fabric blockchain $B$, verifier $V$
\ENSURE Authentication result: 1 if device $D_i$ is authenticated, 0 otherwise

\STATE \textit{Authentication Process}:

\STATE Device $D_i$ retrieves the stored PUF challenges $C_i$ from the blockchain $B$

\STATE Device $D_i$ generates the corresponding PUF responses $R_i \leftarrow P_i(C_i)$

\STATE Device $D_i$ creates a zk-SNARK proof $\pi_i$ proving knowledge of $R_i$ and $sk_i$:
\begin{equation}
\pi_i \leftarrow \text{Prove}(\text{Hash}(pk_i || R_i) = ID_i \land \text{Knowledge}(sk_i))
\end{equation}

\STATE Device $D_i$ sends the proof $\pi_i$ and its identifier $ID_i$ to the verifier $V$

\STATE Verifier $V$ retrieves the device's public key $pk_i$ and PUF challenges $C_i$ from the blockchain $B$

\STATE Verifier $V$ verifies the zk-SNARK proof $\pi_i$ using the public input $(ID_i, pk_i)$:
\begin{equation}
\text{result} \leftarrow \text{Verify}((ID_i, pk_i), \pi_i) = \begin{cases}
1 & \text{if } \pi_i \text{ is a valid proof} \\
0 & \text{otherwise}
\end{cases}
\end{equation}

\RETURN \text{result}
\end{algorithmic}
\end{algorithm}

The Groth16 zk-SNARK scheme used in Algorithm \ref{alg:authentication_protocol} provides several security and performance advantages:
\begin{itemize}
\item \textbf{Zero-knowledge}: The proof $\pi_i$ reveals no information about the PUF responses $R_i$ or the private key $sk_i$, ensuring privacy.
\item \textbf{Succinctness}: The proof size is constant, regardless of the complexity of the statement being proven, enabling efficient transmission and storage.
\item \textbf{Non-interactivity}: The proof can be generated and verified without interaction between the prover and verifier, facilitating asynchronous authentication.
\item \textbf{Soundness}: It is computationally infeasible for an adversary to generate a valid proof for a false statement, preventing impersonation attacks.
\end{itemize}

The ZKP-based authentication protocol ensures secure and privacy-preserving authentication of IoT devices, as stated in the following theorem:

\textbf{Theorem 2.} \textit{The ZKP-based authentication protocol ensures secure and privacy-preserving authentication of IoT devices.}

\textit{Proof sketch.} The security of the authentication protocol relies on the security of the Groth16 zk-SNARK scheme, which guarantees zero-knowledge, soundness, and non-interactivity. The use of PUF-derived identifiers and private keys ensures that only authentic devices can generate valid proofs. The privacy of the device is maintained, as the proof reveals no sensitive information. $\square$

\subsection{Secure Transaction Processing}
The secure transaction processing protocol ensures the integrity and privacy of IoT transactions using zk-SNARKs and smart contracts. The protocol is described in Algorithm \ref{alg:transaction_processing}.

\begin{algorithm}
\caption{Secure Transaction Processing}
\label{alg:transaction_processing}
\begin{algorithmic}[1]
\renewcommand{\algorithmicrequire}{\textbf{Input:}}
\renewcommand{\algorithmicensure}{\textbf{Output:}}
\REQUIRE Authenticated IoT device $D_i$ with identifier $ID_i$ and cryptographic keys $K_i$, Hyperledger Fabric blockchain $B$, smart contract $SC$, transaction $T_i$
\ENSURE Transaction execution result: 1 if the transaction is executed successfully, 0 otherwise

\STATE \textit{Transaction Processing}:

\STATE Device $D_i$ generates a zk-SNARK proof $\pi_{T_i}$ proving the integrity of the transaction data:
\begin{equation}
\pi_{T_i} \leftarrow \text{Prove}(\text{Integrity}(T_i))
\end{equation}

\STATE Device $D_i$ signs the transaction $T_i$ using its private key $sk_i$:
\begin{equation}
\sigma_{T_i} \leftarrow \text{Sign}(T_i, sk_i)
\end{equation}

\STATE Device $D_i$ submits the transaction $T_i$, proof $\pi_{T_i}$, and signature $\sigma_{T_i}$ to the smart contract $SC$

\STATE Smart contract $SC$ verifies the transaction signature $\sigma_{T_i}$ using the device's public key $pk_i$:
\[
\text{sig\_result} \leftarrow \begin{cases}
1 & \text{if } \text{Verify}(T_i, \sigma_{T_i}, pk_i) = 1 \\
0 & \text{otherwise}
\end{cases}
\]

\STATE Smart contract $SC$ verifies the zk-SNARK proof $\pi_{T_i}$ to ensure the integrity of the transaction data:
\[
\text{proof\_result} \leftarrow \begin{cases}
1 & \text{if } \text{Verify}(T_i, \pi_{T_i}) = 1 \\
0 & \text{otherwise}
\end{cases}
\]

\IF{$\text{sig\_result} = 1$ \AND $\text{proof\_result} = 1$}
\STATE Smart contract $SC$ executes the transaction $T_i$ and updates the blockchain state
\RETURN 1
\ELSE
\STATE Transaction $T_i$ is rejected, and the blockchain state remains unchanged
\RETURN 0
\ENDIF
\end{algorithmic}
\end{algorithm}

The zk-SNARK proof $\pi_{T_i}$ in Algorithm \ref{alg:transaction_processing} ensures the integrity of the transaction data without revealing the actual data. The smart contract verifies both the transaction signature and the zk-SNARK proof before executing the transaction, preventing unauthorized modifications and ensuring data privacy.

\textbf{Theorem 3.} \textit{The secure transaction processing protocol ensures the integrity and privacy of IoT transactions.}

\textit{Proof sketch.} The integrity of the transaction data is guaranteed by the zk-SNARK proof $\pi_{T_i}$, which ensures that the data has not been tampered with. The privacy of the transaction is maintained, as the proof reveals no information about the actual data. The use of digital signatures prevents unauthorized modifications and provides non-repudiation. $\square$

\subsection{Security Analysis}
\label{subsec:security_analysis}
The proposed framework provides a comprehensive security solution for blockchain-based IoT systems by leveraging the unique properties of PUFs and ZKPs. In this section, we analyze the security of the framework against various types of attacks and discuss its resistance to common vulnerabilities.

\subsubsection{Device Impersonation Attacks}
The use of PUFs for device identification ensures that each IoT device has a unique and unclonable identifier based on its physical characteristics. An attacker attempting to impersonate a legitimate device would need to physically possess the device and have access to its PUF to generate the correct responses. Moreover, the integration of PUFs with public-key cryptography in the device registration process (Algorithm \ref{alg:device_registration}) binds the device's identity to its public key, making it infeasible for an attacker to impersonate the device without access to the corresponding private key.

\subsubsection{Man-in-the-Middle Attacks}
The ZKP-based authentication protocol (Algorithm \ref{alg:authentication_protocol}) protects against man-in-the-middle attacks by ensuring that the communication between the IoT devices and the blockchain network is secure and authenticated. The use of zk-SNARKs allows devices to prove their authenticity without revealing sensitive information, such as their private keys or PUF responses. An attacker attempting to intercept or manipulate the communication would be unable to generate valid proofs or signatures as they do not possess the required secret information.

\subsubsection{Data Tampering Attacks}
The secure transaction processing workflow (Algorithm \ref{alg:transaction_processing}) ensures the integrity of the transaction data through the use of zk-SNARKs and digital signatures. Each transaction is accompanied by a zk-SNARK proof, which verifies the integrity of the transaction data without revealing the data itself. Additionally, transactions are signed using the device's private key, providing non-repudiation and protecting against unauthorized modifications. An attacker attempting to tamper with the transaction data would need to forge a valid proof and signature, which is computationally infeasible due to the security properties of zk-SNARKs and digital signatures.

\subsubsection{Replay Attacks}
The proposed framework inherently resists replay attacks through the use of unique challenges and proofs for each authentication session and transaction. In the ZKP-based authentication protocol (Algorithm \ref{alg:authentication_protocol}), the device retrieves a fresh set of challenges from the blockchain for each authentication attempt, ensuring that the generated proofs are unique and cannot be reused. Similarly, in the secure transaction processing workflow (Algorithm \ref{alg:transaction_processing}), each transaction is accompanied by a unique zk-SNARK proof, preventing an attacker from replaying previously captured proofs.

\subsubsection{Quantum Resistance}
The security of the proposed framework relies on the underlying cryptographic primitives, such as hash functions, digital signatures, and zk-SNARKs. While quantum computers pose a threat to certain cryptographic algorithms, such as those based on integer factorization (e.g., RSA) or discrete logarithms (e.g., ECDSA), the framework can be adapted to use post-quantum cryptographic primitives to ensure long-term security. For example, hash-based signatures \cite{bernstein_post-quantum_2017} and lattice-based zk-SNARKs \cite{gennaro_lattice-based_2018} can be employed to provide quantum resistance.

\section{Implementation and Results}
\label{sec:implementation_results}

To evaluate the performance and feasibility of our proposed framework, we conducted experiments using a prototype implementation. This section presents the experimental setup, results, and analysis of the framework's performance in terms of time, memory usage, and proof size.

\subsection{Experimental Setup}
\label{subsec:experimental_setup}

The experiments were conducted on a Lenovo ThinkStation P330 Tiny with the following specifications:

\begin{itemize}
\item Memory: 32 GiB
\item CPU: Intel Core i7-8700T @ 2.40GHz
\item Graphics Card: NVIDIA Quadro P1000
\item Operating System: Ubuntu 22.04.4 LTS
\end{itemize}

The prototype implementation was developed using JavaScript and the Circom library for zk-SNARK circuit compilation and proof generation. The experiments were run for 50 iterations to obtain a representative sample of the framework's performance metrics.

\subsection{Off-chain Experiments}
\label{subsec:offchain_experiments}
It is important to note that the experiments conducted in this study were performed off-chain, meaning that the prototype implementation was not deployed on a live Hyperledger Fabric network. The off-chain experiments allowed us to evaluate the performance and feasibility of the proposed framework in a controlled environment, focusing on the computational aspects of the zk-SNARK scheme and the PUF-based authentication protocol. While off-chain experiments provide valuable insights into the framework's performance, future work will involve deploying the framework on a live Hyperledger Fabric network to assess its performance in a realistic blockchain environment and to measure transaction-related metrics.

\subsection{Trust Setup Time}
\label{subsec:trust_setup_time}

The trust setup is a one-time process that generates the necessary parameters for the zk-SNARK scheme used in the proposed framework. The trust setup time is an important metric to consider, as it impacts the initial deployment of the framework.

Our experiments measured the trust setup time to be 1,415.60 ms. This indicates that the trust setup process is relatively efficient and does not introduce a significant overhead during the initial configuration of the framework.

\subsection{Time Metrics}
\label{subsec:time_metrics}

\begin{figure}
\centering
\includegraphics[width=\linewidth]{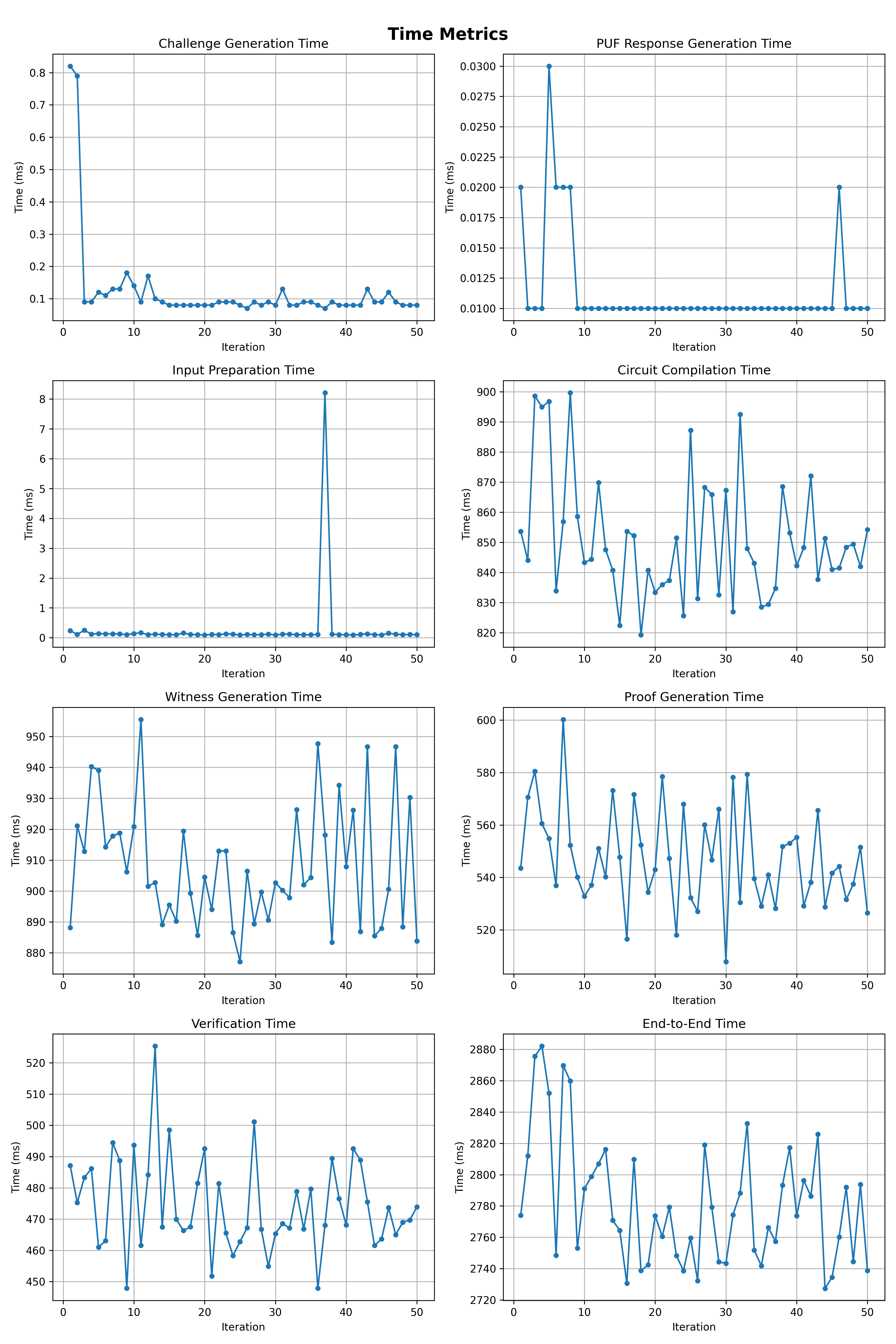}
\caption{Time metrics for the proposed framework across 50 iterations.}
\label{fig:time_metrics}
\end{figure}

Figure \ref{fig:time_metrics} presents the time metrics for various components of the proposed framework across the 50 iterations. The metrics include challenge generation time, PUF response generation time, input preparation time, circuit compilation time, witness generation time, proof generation time, verification time, and end-to-end time.

The results show that the time taken for each component remains relatively consistent across the iterations, with circuit compilation, witness generation, proof generation, and verification being the most time-consuming steps. The end-to-end time, which represents the total time taken for a complete authentication and transaction processing cycle, averages around 2,800 ms.

\subsection{Proof Size}
\label{subsec:proof_size}

\begin{figure}
\centering
\includegraphics[width=0.5\linewidth]{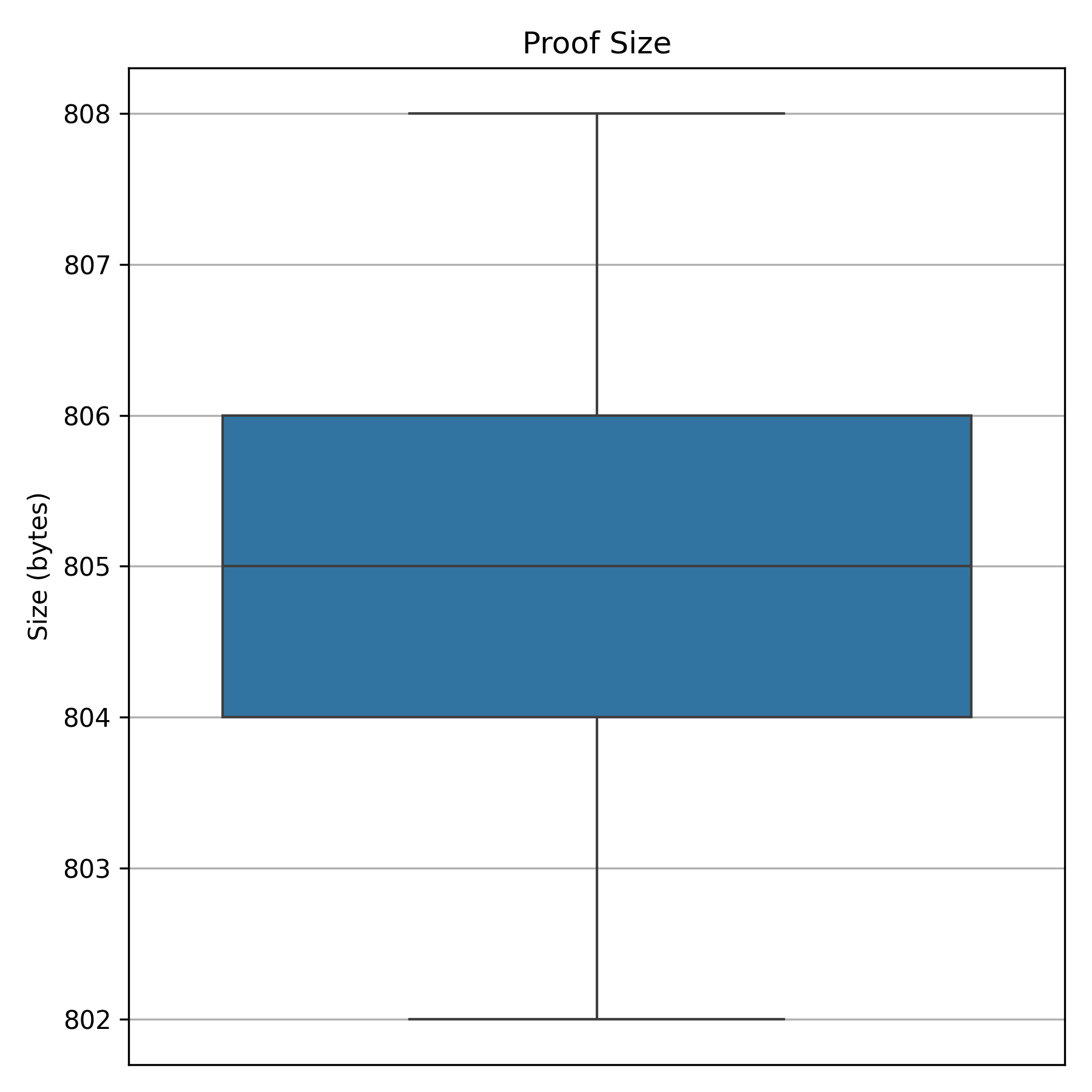}
\caption{Box plot of proof sizes across 50 iterations.}
\label{fig:proof_size_boxplot}
\end{figure}

Figure \ref{fig:proof_size_boxplot} presents a box plot of the proof sizes generated by the framework across the 50 iterations.

The proof sizes remain consistent, with a median size of 805 bytes and minimal variation across the iterations. This demonstrates the compactness of the generated zk-SNARK proofs, which is crucial for efficient storage and transmission in resource-constrained IoT environments.

\subsection{Average Time and Memory Usage Metrics}
\label{subsec:average_metrics}

\begin{figure}
\centering
\includegraphics[width=\linewidth]{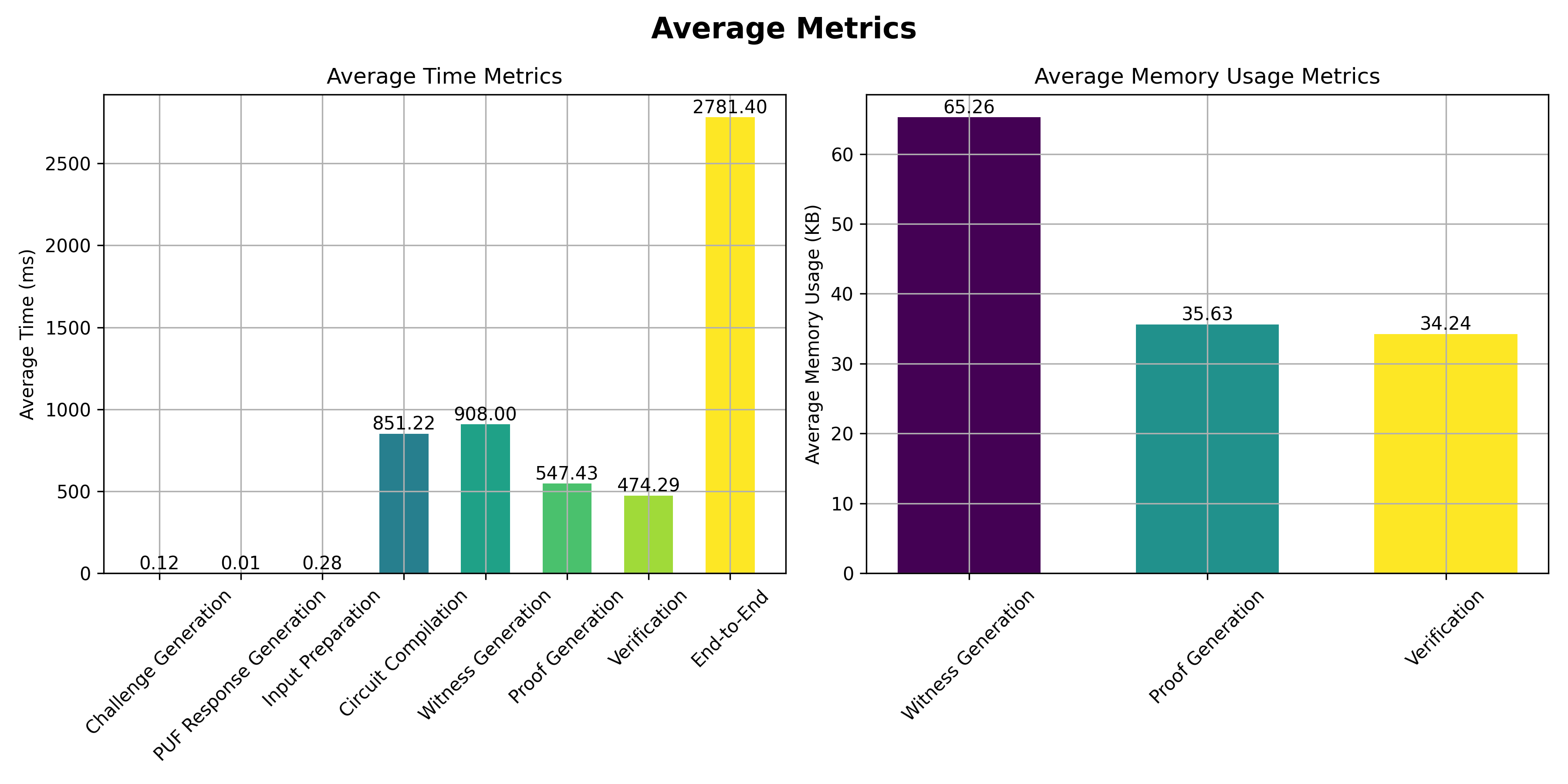}
\caption{Average time and memory usage metrics for the proposed framework.}
\label{fig:average_metrics_barplots}
\end{figure}

Figure \ref{fig:average_metrics_barplots} presents bar plots of the proposed framework's average time and memory usage metrics.

The average time metrics show that circuit compilation, witness generation, proof generation, and verification are the most time-consuming steps, with end-to-end time averaging around 2,800 ms. The average memory usage metrics indicate that witness generation consumes the most memory at approximately 70 KB, followed by proof generation and verification at around 35 KB each.

\section{Discussion and Conclusion}
\label{sec:conclusion}

This paper proposes a framework for securing blockchain-based IoT systems by integrating PUFs and ZKPs within a Hyperledger Fabric environment. The framework addresses the challenges of device authentication, data privacy, and scalability by leveraging the unique properties of PUFs and ZKPs. The proposed system architecture combines IoT devices equipped with PUFs, a Hyperledger Fabric blockchain network, and a ZKP-based authentication and communication protocol.

As mentioned in the experimental setup, the system used for conducting the experiments is significantly more powerful than typical resource-constrained IoT devices. Consequently, the performance metrics obtained from these experiments may not directly reflect the framework's performance on actual IoT devices. However, the proposed framework's use of lightweight cryptographic primitives makes it scalable and adaptable to resource-constrained environments.

To assess the computational complexity and resource requirements of the proposed framework on IoT devices, we provide a theoretical analysis. The zk-SNARK scheme used in the framework has a computational complexity of $\mathcal{O}(n)$ for proof generation, where $n$ is the size of the arithmetic circuit representing the statement being proven, and $\mathcal{O}(1)$ for proof verification. The computational complexity of the PUF-based authentication protocol depends on the specific PUF implementation, with most designs having a complexity of $\mathcal{O}(k)$ for response generation and verification, where $k$ is the size of the challenge-response pairs. The memory requirements for proof generation and verification are constant, as demonstrated by the experimental results. Based on this analysis and the experimental results obtained on the powerful machine, we estimate that the framework can be efficiently deployed on resource-constrained IoT devices with appropriate optimizations and resource management techniques.

In future work, we plan to deploy the proposed framework on a live Hyperledger Fabric network and evaluate its performance in a realistic blockchain environment. Additionally, we will conduct experiments on resource-constrained IoT devices, such as Raspberry Pi, to obtain more representative performance metrics and assess the framework's feasibility in real-world IoT scenarios. Furthermore, we will investigate the integration of our framework with other blockchain platforms, such as Ethereum and Corda, to demonstrate its adaptability and potential for wider adoption.

\bibliographystyle{IEEEtran} 
\bibliography{references} 

\end{document}